\documentclass[structabstract]{aa}
\usepackage{graphicx}
\usepackage{longtable}
\usepackage{lscape}
\usepackage{rotating}
\usepackage{ulem}
\usepackage{txfonts}
\def\arcsec{$^{\prime\prime}$}

\def\NII{N{\sc ii}}

\def\OIII{O{\sc iii}}

\def\epx{~e$^{-1}$\,pixel$^{-1}$}

\begin{document}
   \title{Central Stars of Galactic Planetary Nebulae II
\thanks{Based on data collected at the Complejo Astron\'omico
El Leoncito (CASLEO), which is operated under agreement between the
Consejo Nacional de Investigaciones Cient\'{i}ficas y T\'ecnicas de
la Rep\'ublica Argentina y Universidades Nacionales de La Plata,
C\'ordoba y San Juan, Argentina.}}

\subtitle{New OB-type and emission-line stars}

   \author{W. A. Weidmann
          \inst{1}\fnmsep\thanks{Fellow of Consejo Nacional de Investigaciones
               Cient\'ificas y T\'ecnicas (CONICET), Argentina.}
    \and
          R. Gamen\inst{2}\fnmsep\thanks{Member of Carrera del Investigador CONICET,
          Argentina.}
          }

   \institute{Observatorio Astron\'omico C\'ordoba,
              Universidad Nacional de C\'ordoba, Argentina\\
              \email{walter@mail.oac.uncor.edu}
    \and
             Instituto de Astrof\'isica de La Plata, (CCT La Plata-CONICET, 
             Universidad Nacional de La Plata)\\
             \email{rgamen@fcaglp.unlp.edu.ar}
             }


  \abstract
{There are more than 3000 confirmed and probable known Galactic 
planetary nebulae, but central star spectroscopic information 
is available for only 13\% of them.}  
{We undertook a spectroscopic survey of central stars of PNe 
 to identify their spectral types.}
{We performed spectroscopic observations, at low resolution, with 
the 2-m telescope at CASLEO, Argentina.}
{We present the spectra of 46 central stars of PNe, 
most of them are OB-type and emission-line stars.}
{}

   \keywords{surveys --
                planetary nebulae: general --
                Stars: Wolf-Rayet}

   \maketitle
%

\section{Introduction}
\label{intro}


Planetary nebulae (PNe) are some of the most beautiful objects in the sky, 
there are about 3000 catalogued PNe in our galaxy, but only 13\% of their 
progenitors have been spectroscopically identified. The variety of 
spectral types in these stars was shown by Weidmann \& Gamen 
(\cite{ww3}, hereafter paper I).  

The identification of the ionizing star of a PN is not always an easy 
work; most of them are optically faint objects (low luminosity)
and sometimes they are not at the geometric center, 
because the nebula interacts with the interstellar medium
(Tweedy \& Napiwotzki \cite{intera}). 
Also, PNe are concentrated toward the plane and bulge
of the Galaxy, where crowding and interstellar dust 
make difficult to observe and identify they progenitors.
The fact that the PNe are easily confused with other types of objects 
(Frew \& Parker \cite{fp10}) complicates this picture even more,  and
introduces confusion in statistical works. At present, many non-PNe
remain hidden in the catalogs of PNe (Paron \& Weidmann \cite{ww2}).

A great effort has been made in the recent years to improve and increase 
the spectroscopic observations of central stars of PN (CSPN), which 
together with stellar atmosphere models should help to understand the 
physics of this heterogeneous group of stars.
However, the faintness of most of the CSPNe makes their
spectroscopic observation a very time-demanding  task, 
and thus the studies of these objects progress slowly.
A consequence of all these problems is the very low number of new identified 
OB-type CSPNe (which are harder to detect than emission-line stars).
There were 60  OB  stars listed in Acker et al. (\cite{ack92}), and only 
seven newly identified OB stars have been reported until 2010 (see paper I).

We are carrying out a spectroscopic survey of unclassified CSPN (see Paper I), 
the first results of which are presented in this second paper.
We describe the 
observations and reduction of data in detail (Sect.~\ref{obs}). We analyze 
and show the spectra of OB-type and emission-line stars for classification 
purposes and inter-comparison studies (Sect.~\ref{stcspn}).
We hope that this work will help to guide future observations and lead to
a more reliable determination of stellar parameters of the objects
presented here, 
in particular the absorption-line central stars.


\section{Observations and data reduction}
\label{obs}

The REOSC spectrograph attached
to the 2.15-m telescope at CASLEO, Argentina, 
was employed in this survey.
It is equipped with a CCD Tek 1024$\times$1024
pixels 
(24~$\mu$m) and a read-noise of 7.4\epx.

A 300~line mm$^{-1}$ grating, named \#270 and blazed in the blue,
was used, which yielded a dispersion of 
3.4~\AA\, pixel$^{-1}$
in first order (a resolving power \textit{R} $\thickapprox 2000$).
In some nights, a grating of 600~line mm$^{-1}$ (\#260) was used
(1.6~\AA\, pixel$^{-1}$).
The gratings were used with an unique tilt, giving a typical wavelength
range of 3500--7000~\AA\ (3875-5530~\AA\ with the \#260).
The slit was always centered on the CSPN, oriented 
in the direction East-West, and opened to 
3\arcsec (consistent with the seeing at the site).
The wavelength calibration was performed using Cu-Ne-Ar comparison arcs. 
At least two spectrophotometric standards
(from Hamuy et al., \cite{hamu})
were observed during each observing night, and were used for flux 
calibration
(these stars are detailed in Table~\ref{std}).

The 2D optical spectra were processed using
IRAF\footnote{IRAF: the Image Reduction and Analysis
Facility is distributed by the National Optical
Astronomy Observatories, which is operated by the
Association of Universities for Research in Astronomy,
Inc. (AURA) under cooperative agreement with the
National Science Foundation (NSF).},
following standard techniques.

In most of our spectra we did not subtract 
the nebular emission lines. For that reason some lines were
useless for classification purposes.

The spectra were flux-calibrated with the standard stars 
shown in Table~\ref{std} and the extinction coefficients
calculated for CASLEO by Minniti et al. (\cite{mini}), 
and dereddened using the $c(\mathrm{H}\beta)$
published by Tylenda et al. (\cite{ty92}).

\begin{figure}
   \centering
   \includegraphics[width=9.5cm]{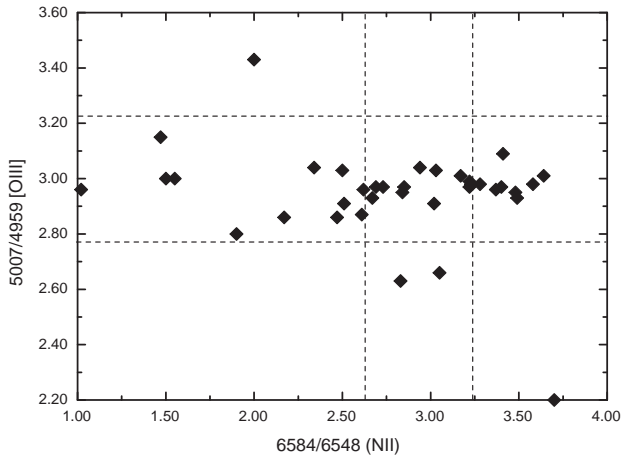}
   \caption[]{The intensity ratios of the \ion{O}{iii} 
             4959, 5007\AA \ and \ion{N}{ii} 6548, 6584\AA \ lines.
             The region defined indicates the value expected
             from the observational data compiled by
             Acker et al. \cite{ack89}.
             }
   \label{line-flux}
\end{figure}

The quality of the data and their reduction can
be assessed by looking at the line ratios of
[\ion{O}{iii}] $5007/4959$~\AA \ and
[\ion{N}{ii}] $6584/6548$ doublets, which
are not very sensitive to nebular conditions.
Figure~\ref{line-flux} shows that the
[\OIII] line ratios are lying around the expected value of
3.01 (Acker et al. \cite{ack89}. The situation is quite different 
for the [\NII] lines, but this is due to the blending of these 
lines with  H$\alpha$ in our spectra.


\section{Spectral classification}

\subsection{Brief description of spectral types observed in CSPN}
\label{stcspn}

CSPN can be divided into two well defined groups, i.e. 
H-rich and H-poor (M\'endez \cite{men91}).
For stars in the former group, hydrogen is the dominant element, 
their more common spectral types are O and O\,f, and,
to a lesser extent, early B (stars with T$_{\mathrm{eff}}<20000$~K
could not ionize the nebula), B$[$e$]$, sdO and
white dwarf of the type DA and 
DAO\footnote{M\'endez (\cite{men91}) included in the type hgO(H) 
the spectral type DA, DAO and sdO.}.

In the H-poor group, the stellar spectrum is almost free 
of hydrogen, while helium and carbon are the most 
prominent elements. Concerning the spectral characteristics, 
we can distinguish two kind of H-poor CSPN. 
Those whose spectrum is dominated by broad and intense emission 
lines, typical of Wolf-Rayet stars [WR], usually of  
[WC] subtype and a few of [WO].
The [WC] stars are dominated by the C {\sc iii} 5696\AA \ line, 
the C {\sc iv} lines being fainter and disappearing 
after the [WC9] class (see M\'endez \cite{MN1982} and \cite{men91}).
The [WO] stars are dominated by the C {\sc iv}  5806\AA \ 
lines (and high excitation O-lines, e.g. O {\sc vi} 3822\AA).
In addition, there are a couple of debatable galactic [WN] CSPN 
(PMR~5 discovered by Morgan et al. (\cite{morgui}) but questioned 
by Todt et al. (\cite{todtb}), PB~8, classified by Todt 
et al. (\cite{todt}) as [WN/WC] and A~48 classified by 
DePew et al. (\cite{depew})  as a possible [WN] or [WN/WC]) but
with a rather high H abundance of 40\%, which might make a spectral 
type Of-WR(H) a preferable description of the spectrum.

A sub-group of CSPN, defined by Tylenda et al. (\cite{ty93},
hereafter TA1993), show emission lines narrower and weaker 
than the [WR] stars; these stars are called wels (weak 
emission-line stars).
Differences between wels and [WC] are subtle, and there 
is no clear criterion to distinguish between both spectral 
types (see e.g. Marcolino \& de Ara\'ujo \cite{MA2003}). 
The most important feature in wels is the
C {\sc iv}  5806\AA\ systematically weaker and narrower 
than for the [WC4-8]. 
Moreover, C {\sc iii} is very weak or absent in canonical wels. 
The blend of C and N ions at 4650\AA\ is also generally detectable 
in wels. Corradi et al. (\cite{corradi2010}) proposed that it comes 
from the irradiated zone in close binary systems.
However, the wels are probably a heterogeneous group of stars and
certainly not all of them are binary systems.
On the other hand, the nebular contamination might lead to 
an incorrect classification, i.e. some Of stars may have 
key absorption lines filled in by the nebular emission 
and be classified as wels instead.

The H-poor CSPN can also display spectra with absorption lines
(sometime dominated by them), named as PG 1159, $[$WC$]$-PG1159, 
O(He)\footnote{see Rauch et al.~(\cite{rau98}).} and DO.

Finally, there is a small group  of stars with
characteristics of the PG 1159 stars, but that also
show hydrogen absorption lines, these objects
are called hybrid-PG1159 stars 
(Napiwotzki \& Sch\"onberner~\cite{napi91}).

It is important to mention  that some 
spectral types also found in the literature
are currently included in the classification 
scheme described above, for example:
Of-WR(C) into [WC]-PG1159, O(C) into PG 1159 and
Of-WR(H) into wels.


\subsection{Classification and criteria}
\label{crito}

We have tried, whenever possible, to determine the spectral types of 
our CSPN sample. For all those showing a stellar continuum
with H or He absorption lines, we tried to carry out a spectral 
classification in the MK system.
This was done taking the spectral types from 
the Atlas of Walborn \& Fitzpatrick (\cite{nola}). 
For PNe with H-poor cores (i.e., with [WR]-type 
central stars), we followed the criteria defined by TA1993 
and Acker \& Neiner (\cite{an03}, Table~2).
This classification is based on the FWHM and dereddened strength, 
with respect to the continuum, of several emission lines.
Concerning wels, as stated above, there is no robust quantitative 
criterion leading to any certainty in the classification.  
However, we used the work of TA1993 (their Table~6), which is
based on dereddened lines at $\lambda$ 4650\AA\ and \ion{C}{iv} 5806\AA.

\begin{figure*}
   \centering
   \includegraphics[width=0.87\textwidth]{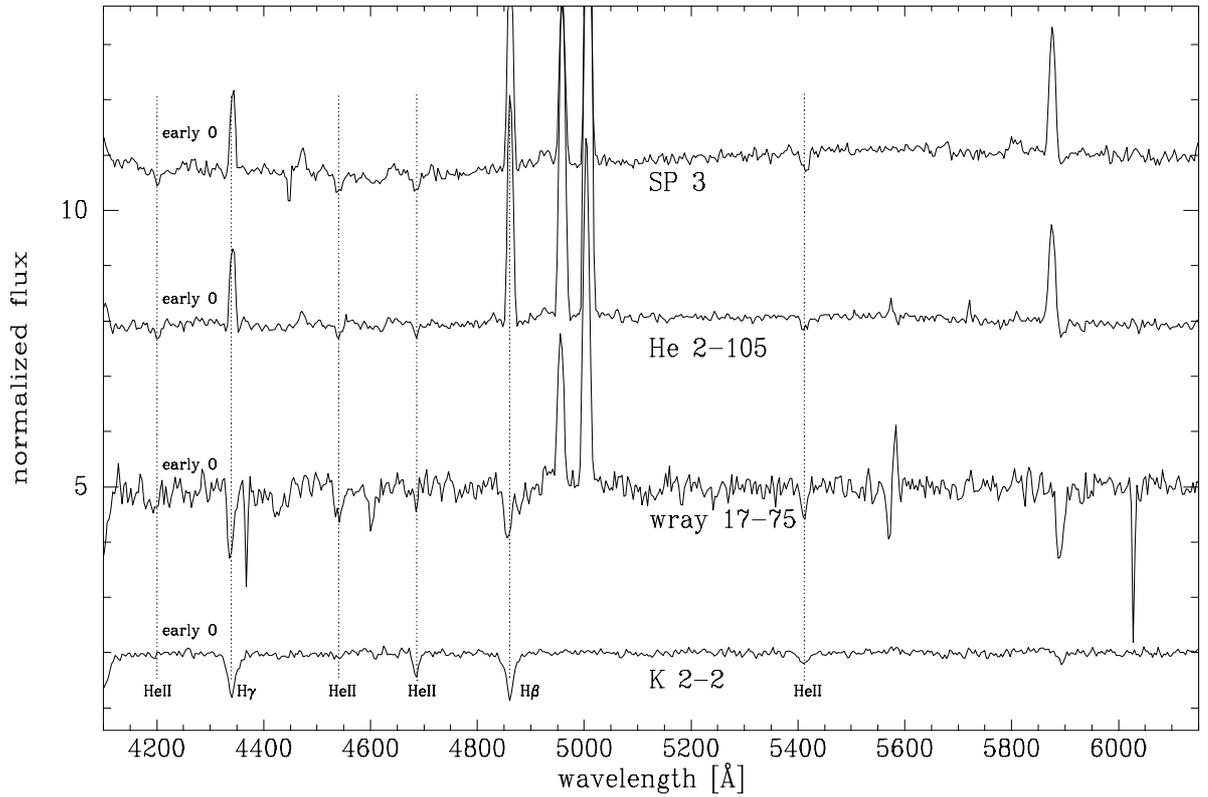}
      \caption[]{Normalized spectra of early-O CSPN 
                 (see Table~\ref{sample}) grouped 
                 according to their spectral classification.
                 The interstellar absorption bands ($\lambda$4428; 
                 $\lambda$5780 and $\lambda$5893) are not indicated.
                 The most important spectral features (absorption or emission) 
                 identified are:
   \ion{He}{ii} $\lambda$4200; H$\gamma$;   \ion{He}{ii} $\lambda$4541;
   \ion{He}{ii} $\lambda$4686; H$\beta$; and \ion{He}{ii} $\lambda$5412.}
         \label{ST-type}
   \end{figure*}

\begin{figure*}
   \centering
   \includegraphics[width=0.87\textwidth]{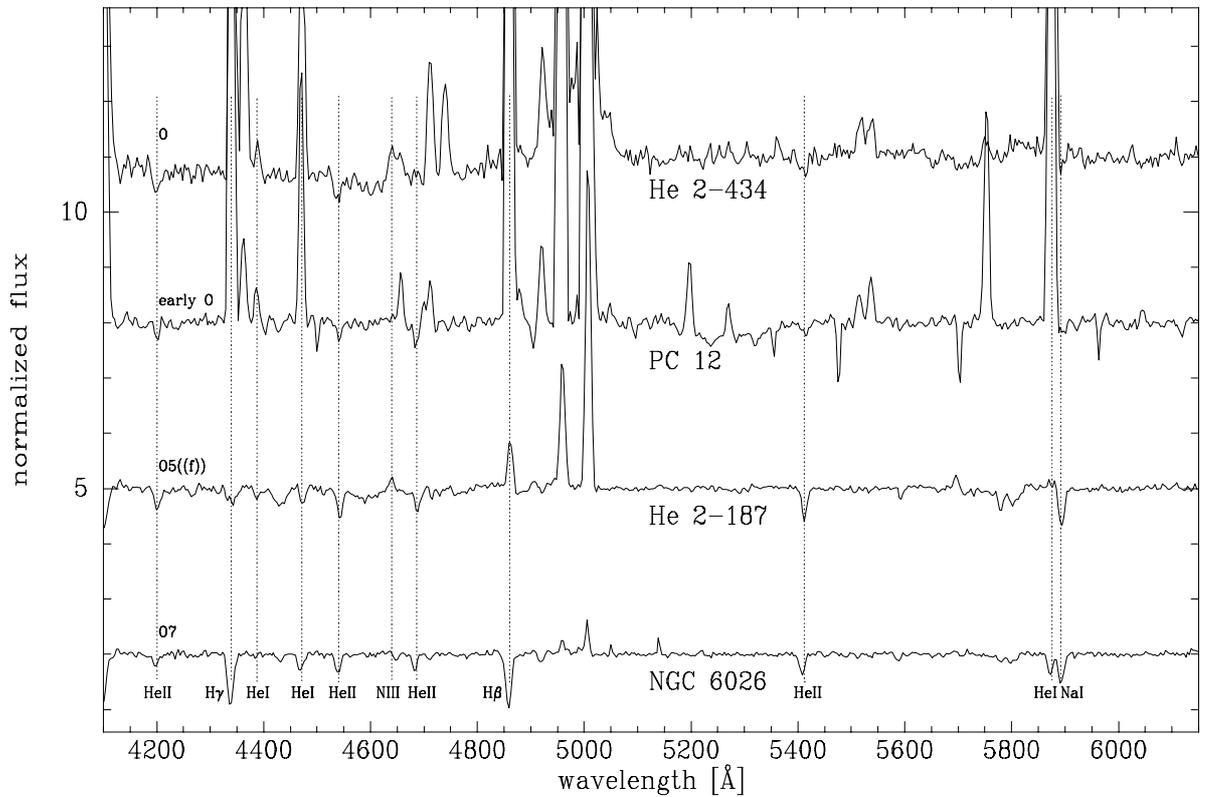}
      \caption[]{Normalized spectra of O-type CSPN. As in Fig.~\ref{ST-type}, 
                 most important spectral lines are identified, plus 
                 \ion{He}{i} lines at 4387, 4541 and 5876\AA.
                 In addition the interstellar line of NaI is indicated.
                 In the spectra of NGC 6026, nebular emission was removed.} 
         \label{ST-type1}
   \end{figure*}

\begin{figure*}
   \centering
   \includegraphics[width=0.87\textwidth]{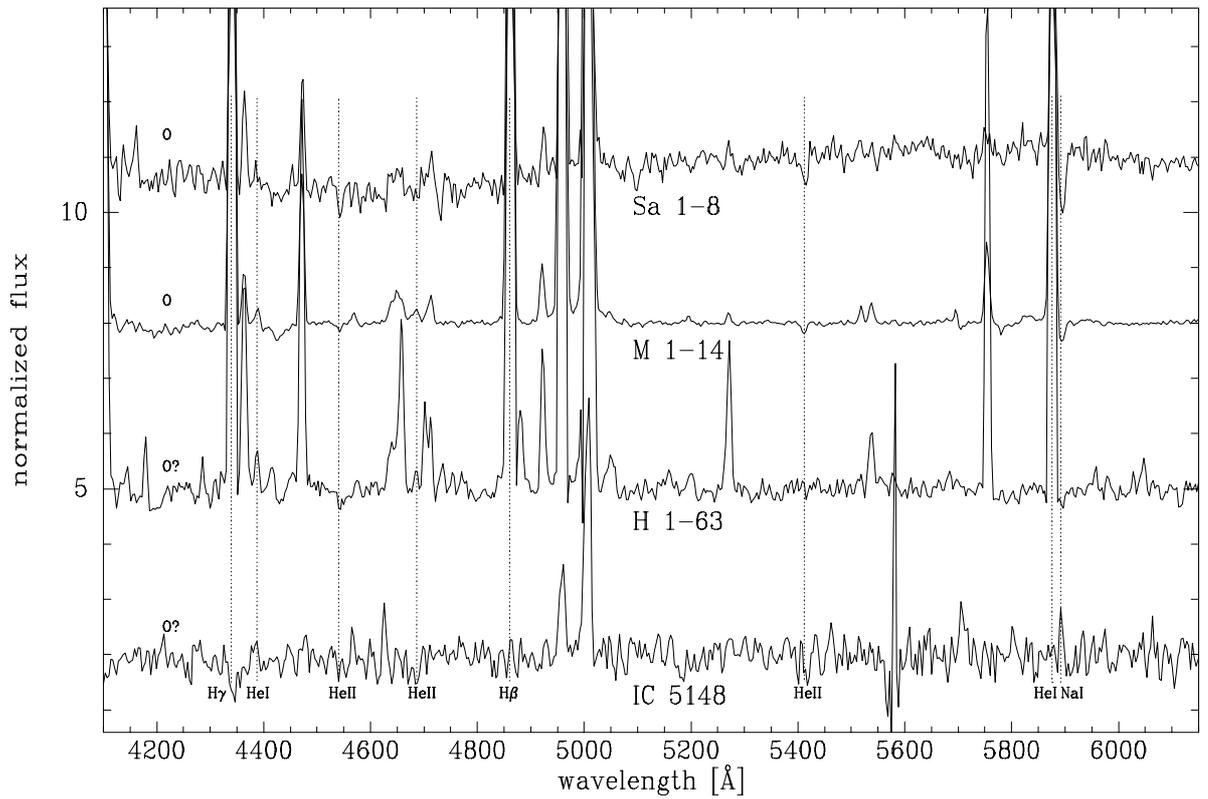}
      \caption[]{Normalized spectra of O-type CSPN. Absorption lines of 
                 \ion{He}{ii} are observed in the spectra. The 
                 interstellar line of NaI is indicated.}
         \label{ST-type2}
   \end{figure*}

\begin{figure*}
   \centering
   \includegraphics[width=0.87\textwidth]{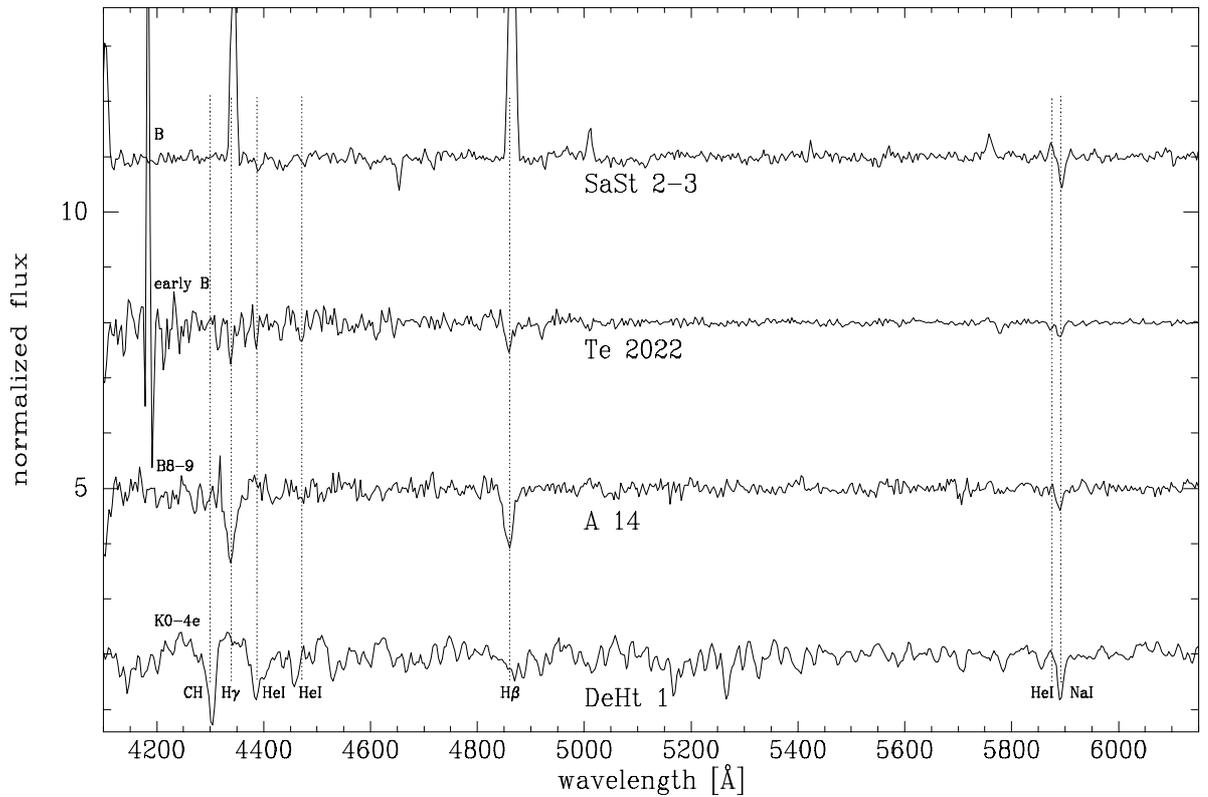}
      \caption[]{Normalized spectra of B-type CSPN, plus DeHt 1. 
                 \ion{He}{ii} is not observed, but 
                 \ion{He}{i}  are noticeable, suggesting a B type.
                 Note absorption lines of the
                 CH G-band $\lambda$4300 and \ion{Na}{i} $\lambda$5892 in DeHt~1.}
         \label{ST-type4}
   \end{figure*}

\begin{figure*}
   \centering
   \includegraphics[width=0.87\textwidth]{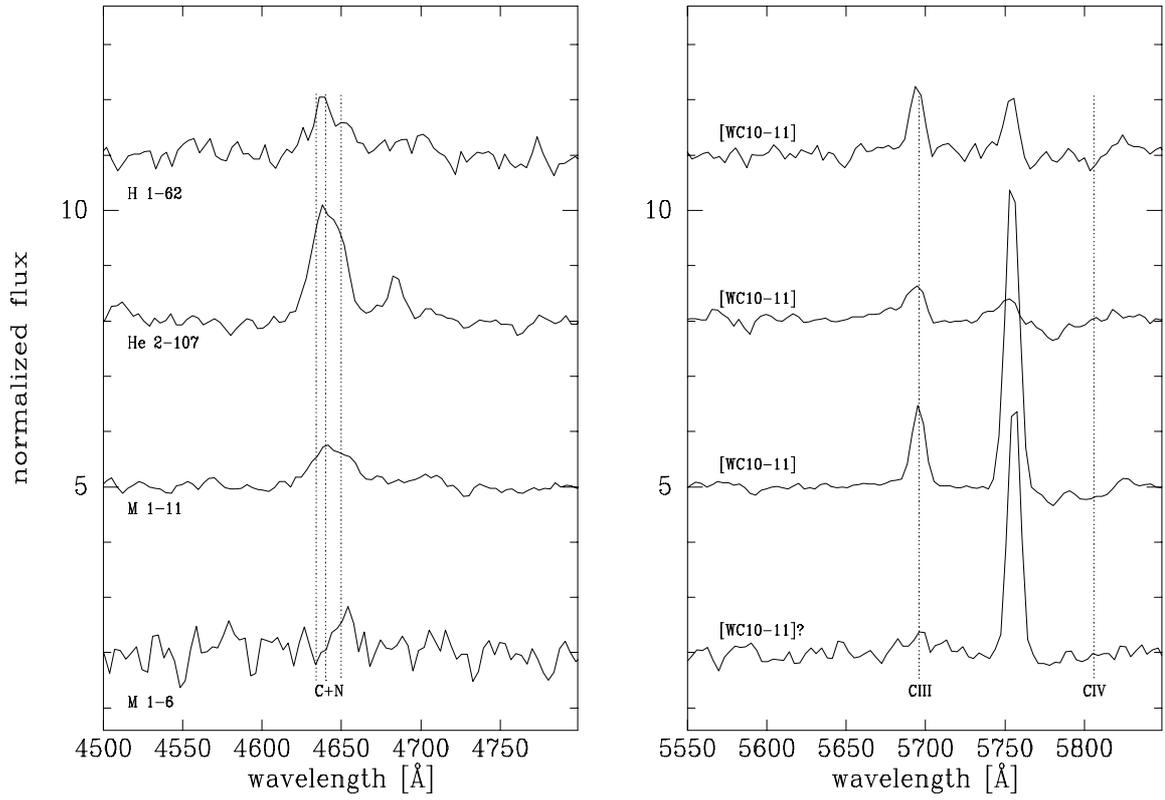}
      \caption[]{Normalized spectra of [WC] CSPN. 
                 The three lines indicated by C+N are \ion{N}{iii} $\lambda$4634, 
                 \ion{N}{iii} $\lambda$4640 and \ion{C}{iii} $\lambda$4650.
                 Note the presence of the emission line \ion{C}{iii} $\lambda$5696 
                 and the absence of \ion{C}{iv} $\lambda$5806.}
         \label{ST-type8}
   \end{figure*}

\begin{figure*}
   \centering
   \includegraphics[width=0.87\textwidth]{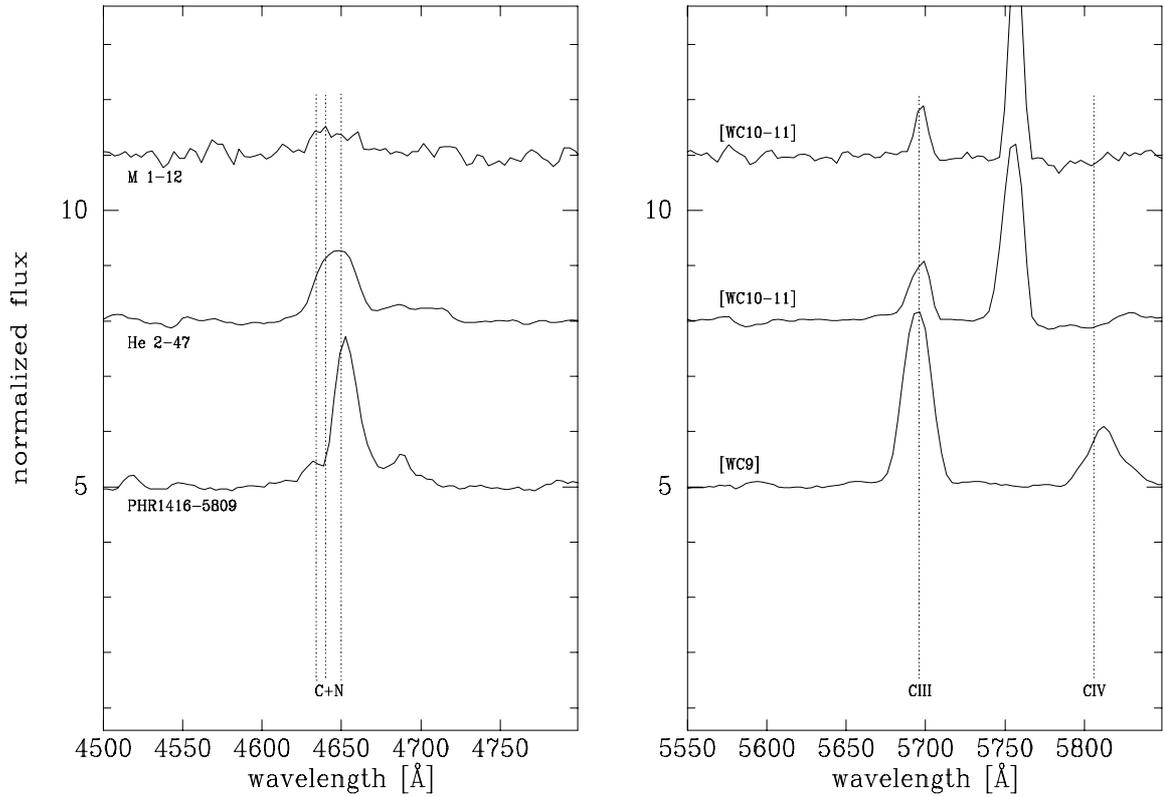}
      \caption[]{Normalized spectra of [WC] CSPN.}
         \label{ST-type7}
   \end{figure*}

\begin{figure*}
   \centering
   \includegraphics[width=0.87\textwidth]{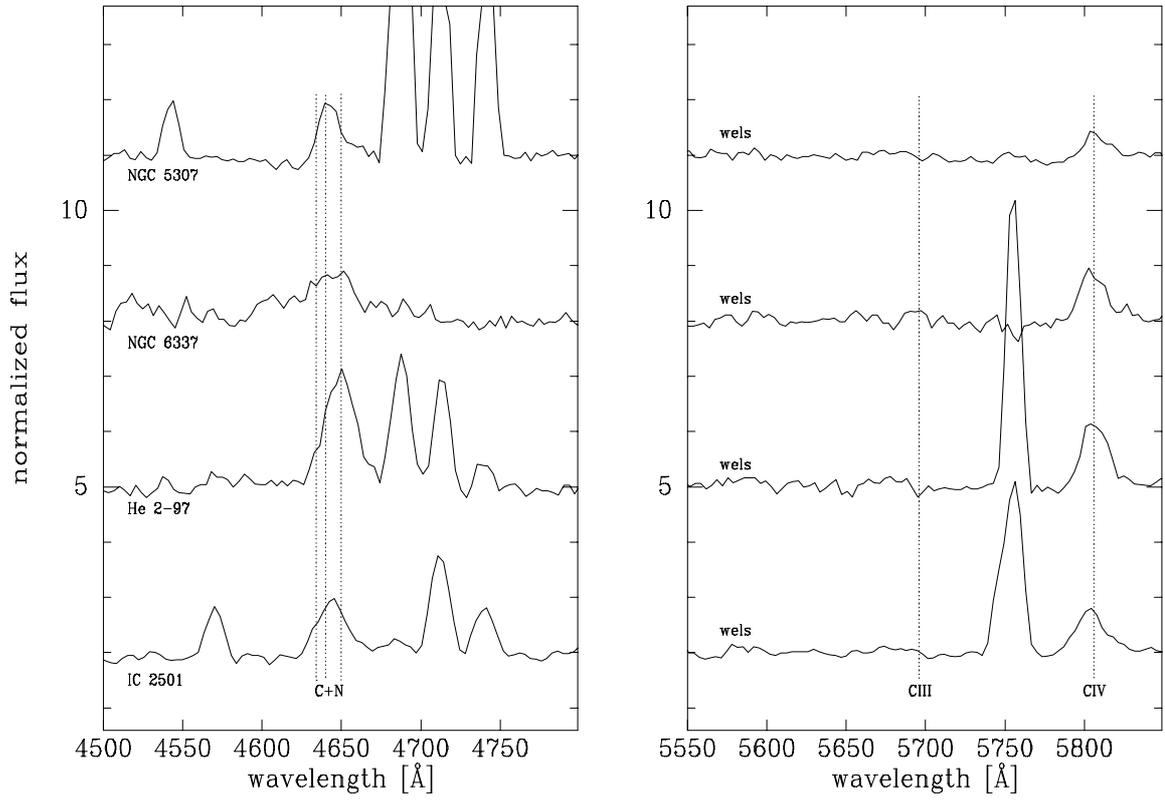}
      \caption[]{Normalized spectra of wels CSPN.
                 Note the absence of \ion{C}{iii} at 5696\AA \ and
                 narrow emission of \ion{C}{iv}.}
         \label{ST-type6}
   \end{figure*}

\begin{figure*}
   \centering
   \includegraphics[width=0.87\textwidth]{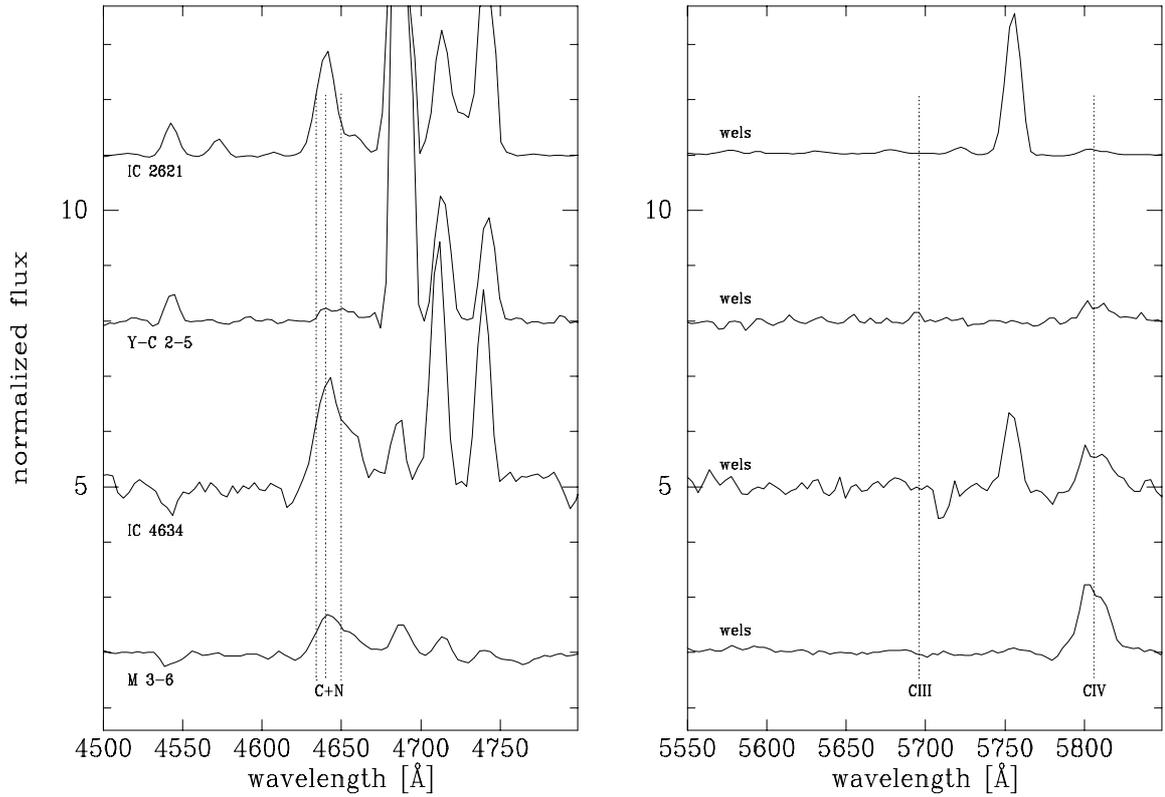}
      \caption[]{Normalized spectra of wels CSPN. 
                Note that \ion{He}{ii} $\lambda$4541 is in absorption 
                in the spectra of IC~4634 and M~3-6.} 
         \label{ST-type10}
   \end{figure*}

\begin{figure*}
   \centering
   \includegraphics[width=0.87\textwidth]{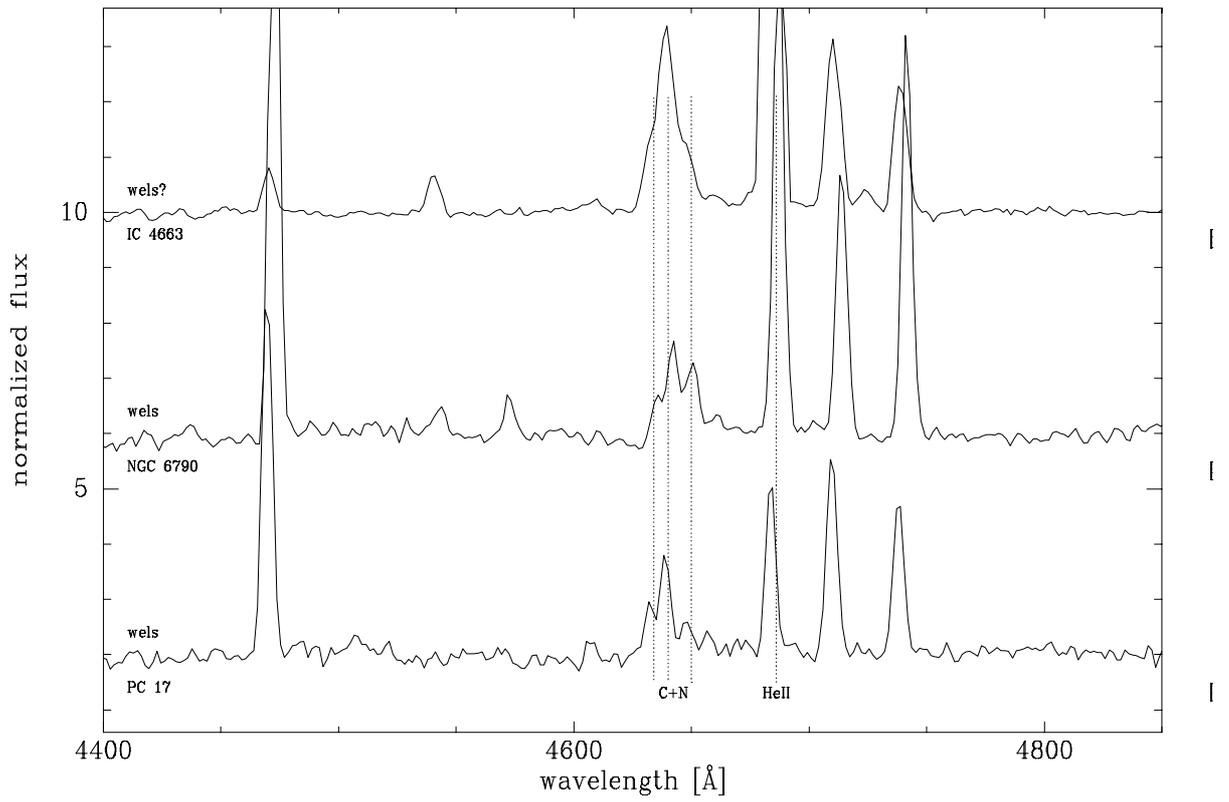}
      \caption[]{Normalized spectra of wels CSPN, but observed 
                with the 600 line~mm$^{-1}$ grating. Note the 
                weak emission of \ion{C}{iv} at 4658\AA.}
         \label{ST-type9}
   \end{figure*}

\begin{figure*}
   \centering
   \includegraphics[width=0.87\textwidth]{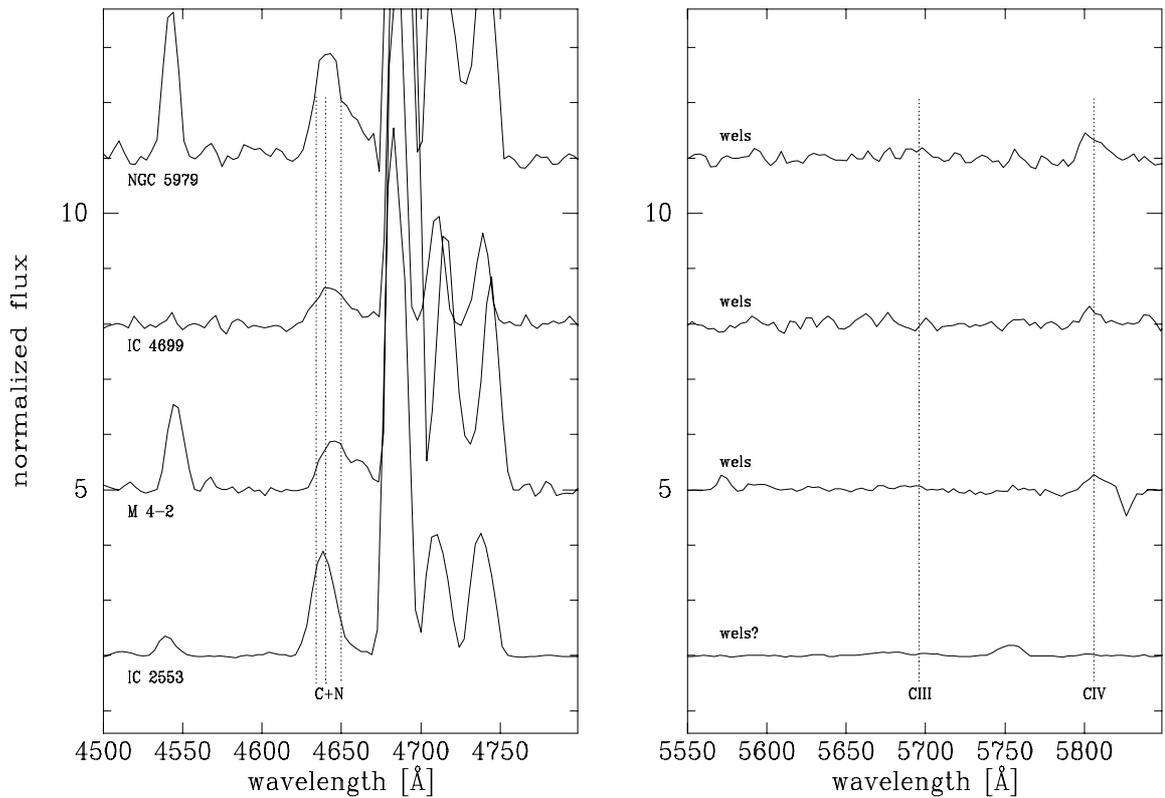}
      \caption[]{Normalized spectra of wels CSPN.}
         \label{ST-type12}
   \end{figure*}

\begin{figure*}
   \centering
   \includegraphics[width=0.87\textwidth]{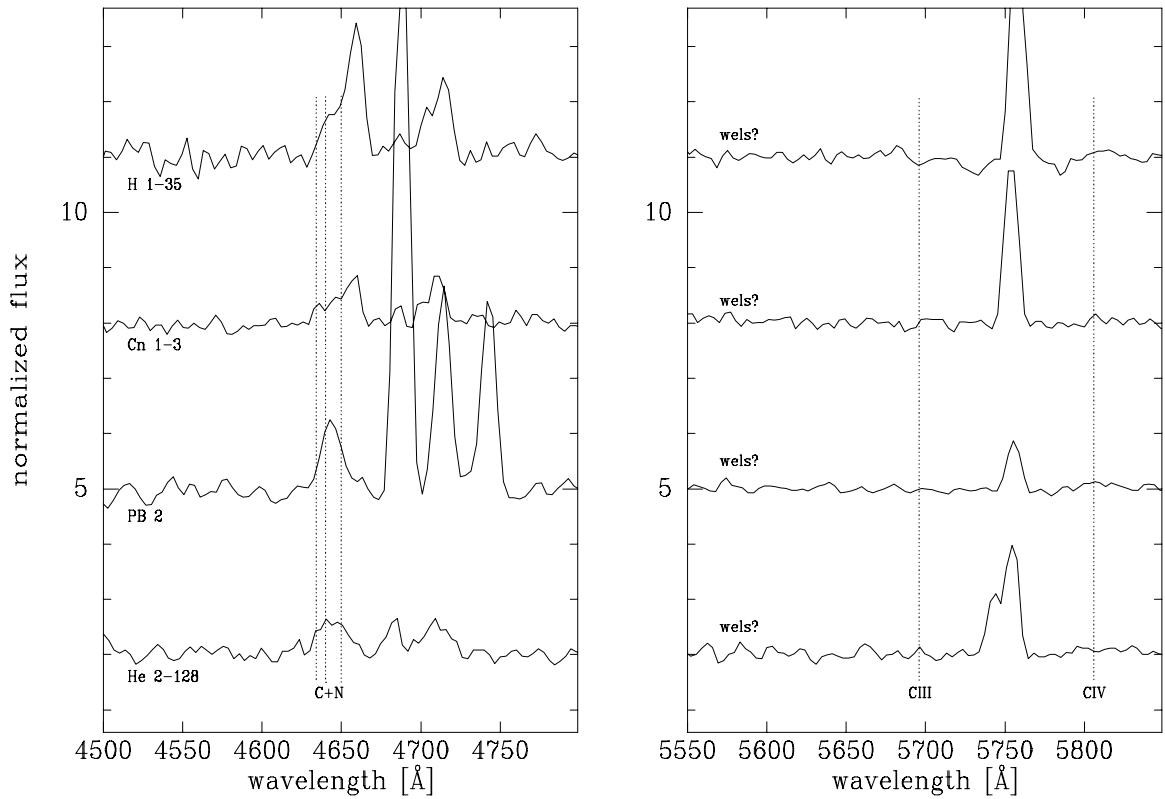}
      \caption[]{Normalized spectra of ``wels?'' CSPN. 
                The unique noticeable feature is the emission at 4650\AA.
	        This emission in the spectrum of H 1-35 and CN 1-3 seems 
                to be dominated by \ion{C}{iv}~4658\AA.}
         \label{ST-type13}
   \end{figure*}

\begin{figure*}
   \centering
   \includegraphics[width=0.87\textwidth]{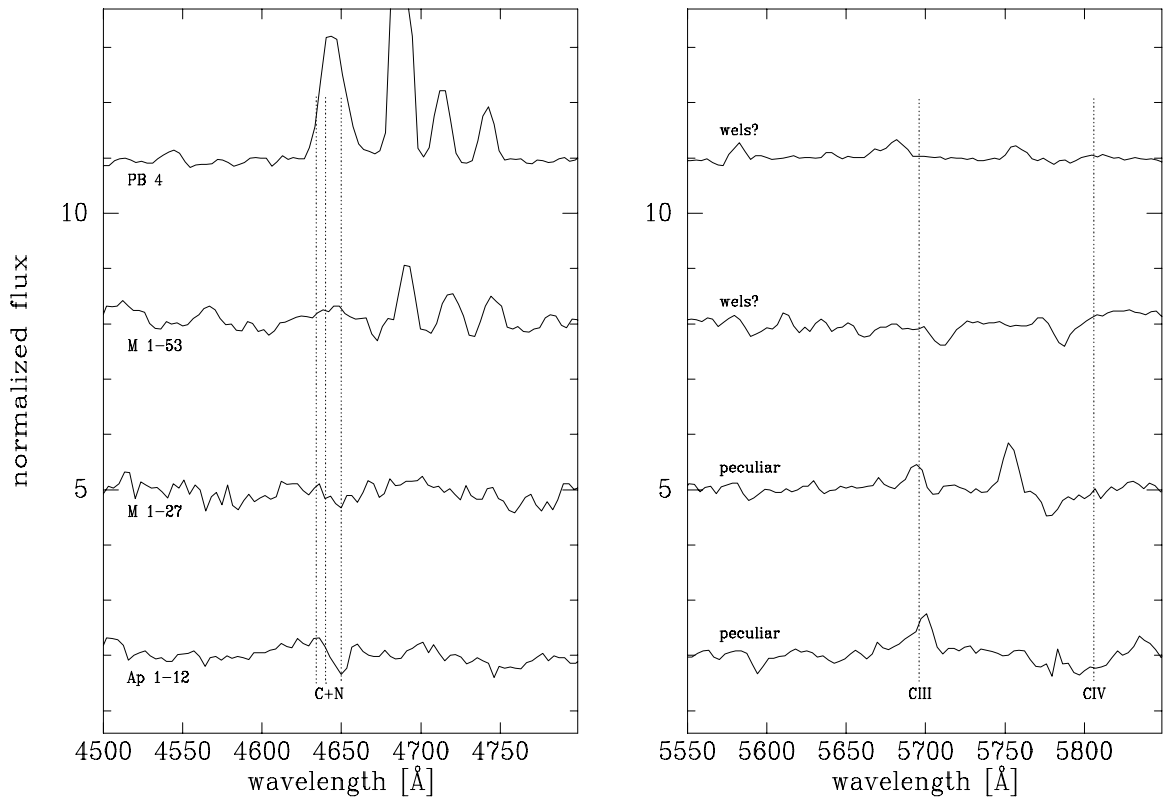}
      \caption[]{Normalized spectra of other CSPN.}
         \label{ST-type11}
   \end{figure*}

Our spectral classifications are presented in the third 
column of Table~\ref{sample}.
The FWHM and W$_\lambda$ of the emission lines used to derive 
those classifications are listed in Table~\ref{emiss1}.
The spectra of the classified CSPN are shown in 
figures~\ref{ST-type}-\ref{ST-type11} (sorted by spectral types).
In the case of emission-line stars, we preferred to show 
only the spectral range where the more intense lines are located.

For those spectra  where quantitative classification was not 
possible, the presence of \ion{C}{iv} $\lambda$5806 and the 
absence of \ion{C}{iii} $\lambda$5696 resulted in a ``wels'' 
type. The detection of the \ion{C}{iii} line indicates a [WC]. 
Finally, 8 CSPN were tagged as ``wels?'' because the lines 
of \ion{C}{iv} 5806\AA\ and \ion{C}{iii} 5696\AA\ are 
absent (or very weak) but we detected the complex at 
4650 \AA. In all these cases we could identify \ion{C}{iv} 
4658\AA\ and \ion{He}{ii} 4686\AA\ (some wels do not have 
emission at \ion{C}{iv}; see Marcolino \& de Ara\'ujo \cite{MA2003}).

The equivalent width and FWHM measured in our spectra are 
quite lower than those published by TA1993, 
perhaps because of the low S/N of our spectra. However, 
the FWHM of the nebular emission line is $\sim$9\AA, 
indicating that the lines shown in Table~\ref{emiss1} 
belong in fact to the stars.
On the other hand, taking into account the mean wavelength 
of the blend at 4650\AA \ we could confirm that, in the wels, 
the principal contributing element is N, as was shown by TA1993.


\subsection{Notes of some individual CSPN}
\label{spectrum}

\textbf{IC~4634}~(PN G000.3+12.2): wels. Emission lines of
\ion{C}{iv} at 5801-11\AA\ seem to be resolved, 
suggesting narrow lines.
Also, the emission blend \ion{C}{iii}+\ion{C}{iv} at 
4654\AA \ is present. 
In addition, we observe absorption lines of \ion{He}{ii} at 
4541\AA. Although we cannot rule out an Of spectral type, 
since a weak emission at 5806 can be observed in very early 
O-type stars, e.g. in the spectrum of HD~93129A (private 
spectrum of Gamen), the strong emissions of \ion{C}{iv} 
suggest a wels rather than an Of star. This object was 
previously classified as continuous by Hyung et al. 
(\cite{hyu}) and Feibelman (\cite{feibe}).

\textbf{IC~5148}~(PN G002.7$-$52.4): O?. We were able to 
minimize the nebula contamination and to identify the
absorption lines of the Balmer series and \ion{He}{ii} 4686\AA\ 
(the later confirming an O-type spectrum).
M\'endez~\cite{men91} classified this star as hgO(H).
Nevertheless IC~5148 does not seem to be an evolved object.

\textbf{Ap~1-12}~(PN G003.3$-$04.6): peculiar.
We identify absorption lines of 
\ion{He}{i} $\lambda\lambda$ 4387, 4471, 4920 and 5876\AA,
\ion{He}{ii} 5412\AA \ (weak), 
\ion{C}{iii} 4650\AA \ and \ion{C}{iv} 5806. 
Moreover, the  emission line of 
\ion{C}{iii} 5696\AA \ is noticeable,
and there is not evidence of \ion{C}{ii}.
G\'orny et al.~(\cite{gorn}) classified this star as a [WCL], 
based on the detection of \ion{C}{iii} 5696\AA. 
Ap~1-12 is spectroscopically similar 
(except by the lack of \ion{C}{ii})
to the peculiar CSPN IRAS~21282+5050
(Acker \& Neiner~\cite{an03}; Crowther et al.~\cite{crow}).

\textbf{A~14}~(PN G197.8$-$03.3): B8-9. In this spectrum 
the Balmer series is clearly visible in absorption. Along 
with the lack of \ion{Ca}{ii} 3933 \AA\ and He {\sc i} 
absorption lines, this seems to indicate a late B spectral type.
This star was classified as B5 III-V by Lutz \& Kaler (\cite{luka}), 
according its dereddened colors.
Such a cool spectral type cannot explain the high ionization of the 
nebula (Bohigas~\cite{bohi}), so it must be a binary system.

\textbf{K~2-2}~(PN G204.1+04.7): early O. We note the presence of the
Balmer lines and strong \ion{He}{ii} lines at 4686\AA\ and 5412\AA, 
whereas there is no sign of \ion{He}{i} lines. 
Therefore, we propose an early-O type for this star. 
Napiwotzki \& Sch\"onberner (\cite{nasch}) classified it as hgO(H).

\textbf{M~1-6}~(PN G211.2$-$03.5), 
\textbf{M~1-11}~(PN G232.8$-$04.7) and 
\textbf{M~1-12}~(PN G235.3$-$03.9):
The emission-line nature of these stars was first 
noted by Kondrat'eva~(\cite{kond}).
However, she did not classified the stars.

\textbf{DeHt~1}~(PN G228.2$-$22.1): K0-4e. 
The most important feature is the very strong  
H$\alpha$ emission. We have taken several echelle 
spectra of this CSPN and observed that the 
emission line is double and its radial velocity is 
variable. While Bond et al. (\cite{bond}) did not 
mention H$\alpha$ emission, they commented on the 
strong \ion{Mg}{ii} and \ion{Ca}{ii} emission, which 
could arise from a fast-rotating cool component 
in a binary system. We do not detect the H and K 
lines in emission; thus they may be variable.

\textbf{SaSt~2-3}~(PN G232.0+05.7): B. Absorption lines 
\ion{He}{i} 4920\AA\ and 4713\AA (weak) and 
\ion{C}{iii} 4650\AA\ (strong) are observed. 
There are also hints of \ion{Si}{iv} 4089\AA\ and 4116\AA.
Kohoutek (\cite{koho97}) suggested this could possibly be a PPN. 
Pereira \& Miranda~(\cite{pemi}) noted He absorption lines and an 
absorption line that they suggested was due to \ion{C}{iv} at 4658.
Stellar absorption lines are also seen in the spectrum 
presented by Dopita \& Hua~(\cite{dohu}).
This object deserves more detailed studies.

\textbf{M~3-6}~(PN G253.9+05.7): wels. \ion{He}{ii} absorption 
lines are observed at 4200 and 4541\AA \ and possibly 5412\AA. 
A strong emission of \ion{C}{iv} 5806\AA\ is seen, as well as 
the typical group of wels emission lines around 4650\AA.
But, as in the case of IC~4634, an Of classification cannot be rejected.
This object was also classified as wels by Acker \& Neiner (\cite{an03}).
Finally, a binary nature could explain this composite spectrum.

\textbf{PHR~1416-5809}~(PN G313.9+02.8): [WC9]. 
A very recent study by DePew et al.~(\cite{depew}) 
also classifies this star as [WC9].

\textbf{NGC~6026}~(PN G341.6+13.7): O7. 
We observe absorption lines of the Balmer series, 
\ion{He}{i} is observed at 4471, 4920, 5876 and 6678\AA, 
and \ion{He}{ii} at 5412, 4686, 4541, 4200 and 4026\AA. 
Possible absorption of \ion{C}{iii} is seen at 4650\AA. 
The intensities of the lines 4541 and 4471 are 
comparable, suggesting a type O7. 
Hillwig et al.~(\cite{hill}), showed that this is a close binary.

\textbf{Sp~3}~(PN G342.5$-$14.3): early O. 
Balmer lines and \ion{He}{ii} 4200\AA, 4541\AA, 
4686\AA, and 5411\AA \ are identified but no 
\ion{He}{i} is present in our spectrum. 
An emission of \ion{C}{iv} 5806\AA \ is present
(see description of M~3-6).
Through observations in the UV, Gauba et al. (\cite{gau}) 
classified it as O3V and possible binary.

\textbf{IC~4699}~(PN G348.0$-$13.8): wels.
Gorny et al.~(\cite{gorn}) had suggested it to 
be a wels by examination of measured lines 
tabulated by Wang \& Liu (\cite{wali}).

\textbf{NGC~6337}~(PN G349.3$-$01.1): wels. 
The spectrum presents an intense emission of \ion{C}{iv} 
at 5806\AA, and another wide emission at 4642\AA.  
Hillwig et al.~(\cite{hill}), showed that this is 
a close binary. In this case, the wels features 
may come from an irradiated secondary.

\textbf{M~1-27}~(PNG356.5$-$02.3): peculiar.
This is a very interesting object; we observed absorption 
lines of \ion{O}{iii} at 5592 and \ion{He}{ii} at 5412\AA. 
But there is a clear emission line of \ion{C}{iii} 
at 5696\AA,  and a lack of \ion{C}{ii}.
On the other hand, R.H.M\'endez  (private communication)
reports other absorption lines, as \ion{C}{iv} $\lambda$5806,
\ion{He}{ii} $\lambda$4686 and H$\gamma$.
Such characteristics are observed in the nucleus of the PNe
K~2-16, IRAS~21282+5050 and Ap~1-12 (see description of this spectrum).
G\'orny et al.~\cite{gorn04} classified it as a [WC11]?


\section{Summary}
\label{con}

We are carrying out a spectroscopic survey of PNe from 
a 2.15-m telescope at CASLEO, Argentina. 
We have performed a quantitative and qualitative
determination of the spectral types of 46 of their 
central stars 
(plus 6 CSPNe classified as ``continuous'' in paper I),
more of them previously unclassified.
Note that 16 of them are OB-type stars, thus
substantially increasing their number among CSPN.
Also, in a few CSPN we have reported some features which
could indicate their binary nature, e.g. DeHt 1.  
Spectroscopic observations at higher spectral resolution 
are required to obtain more accurate classifications.
In particular, the central stars of Ap~1-12 and M~1-27, 
whose spectra are similar to those of the rare objects 
K~2-16 and IRAS~21282+5050, which do not have a reliable 
classification.

With this paper, we complete the description  
of the sample of CSPN presented in Paper I.
We hope that the spectroscopic data presented here 
will provide a guide to future observations, thus
contributing to the better understanding
of the final stages of stellar evolution.


\begin{acknowledgements}

We thank the anonymous referee whose very useful remarks 
helped us to substantially improve this paper.
The CCD and data acquisition system at CASLEO has been financed
by R. M. Rich trough U. S. NSF grant AST-90-15827. This work
has been partially supported by Consejo Nacional de Investigaciones
Cientif\'{i}cas y T\'ecnicas de la Rep\'ublica Argentina (CONICET).
This research has made use of the SIMBAD database,
operated at CDS, Strasbourg, France.

\end{acknowledgements}



\begin{table*}
\caption{Spectral types from our observations. }
\label{sample}
\begin{tabular}{lcc}
\hline\hline\noalign{\smallskip}
 name & PN G & S.T.     \\
\noalign{\smallskip}\hline\noalign{\smallskip}
H 1-62    &  000.0$-$06.8  &  [WC10-11]     \\
PC 12     &  000.1+17.2  &  early O       \\
IC 4634   &  000.3+12.2  &  wels          \\
H 1-63    &  002.2$-$06.3  &  O?            \\
IC 5148   &  002.7$-$52.4  &  O?            \\
Ap 1-12   &  003.3$-$04.6  &  peculiar      \\
M 1-53    &  015.4$-$04.5  &  wels?         \\
Sa 1-8    &  020.7$-$05.9  &   O            \\
NGC 6790  &  037.8$-$06.3  &  wels          \\
A 14      &  197.8$-$03.3  &  B8-9          \\
K 2-2     &  200.1+04.7  &  early O       \\
M 1-6     &  211.2$-$03.5  & [WC10-11]?     \\
DeHt 1    &  228.2$-$22.1  & K0-4e         \\
SaSt 2-3$^*$ &  232.0+05.7  &  B          \\
M 1-11    &  232.8$-$04.7  &  [WC10-11]     \\
M 1-14    &  234.9$-$01.4  &  O             \\
M 1-12    &  235.3$-$03.9  &  [WC10-11]     \\
Y-C 2-5   &  240.3+07.0  &  wels          \\
M 4-2     &  248.8$-$08.5  &  wels          \\
M 3-6     &  253.9+05.7  &  wels          \\
PB 2      &  263.0$-$05.5  &  wels?         \\
PB 4      &  275.0$-$04.1  &  wels?         \\
IC 2501   &  281.0$-$05.6  &  wels          \\
IC 2553   & 285.4$-$05.3 & wels?            \\
He 2-47   & 285.6$-$02.7 &  [WC10-11]       \\
IC 2621   & 291.6$-$04.8 &  wels            \\
He 2-97   & 307.2$-$09.0  &  wels           \\
He 2-105  &  308.6$-$12.2 & early O         \\
NGC 5307  &  312.3+10.5  &  wels          \\
He 2-107  &  312.6$-$01.8  &  [WC10-11]     \\
PHR1416-5809 & 313.9+02.8 & [WC9]        \\
He 2-434  &  320.3$-$28.8  &  O            \\
NGC 5979  &  322.5$-$05.2  & wels          \\
He 2-128  &  325.8+04.5 &  wels?         \\
WRAY 17-75 &  329.5$-$02.2 & early O       \\
He 2-187$^*$ & 337.5$-$05.1 & O5 ((f))    \\
NGC 6026  &  341.6+13.7  &  O7           \\
Sp 3      &  342.5$-$14.3  &  early O     \\
PC 17     &  343.5$-$07.8  &  wels        \\
Cn 1-3    &  345.0$-$04.9  &  wels?       \\
IC 4663   &  346.2$-$08.2  &  wels?       \\
IC 4699   &  348.0$-$13.8  & wels        \\
NGC 6337  & 349.3$-$01.1 & wels          \\
H 1-35    &  355.7$-$03.5  &  wels?       \\
M 1-27$^*$ & 356.5$-$02.3  & peculiar     \\
Te 2022   &  358.8+00.0  &  early B    \\
\hline
\end{tabular}
\tablefoot{The PN are denoted by their common name and by 
           their PN G designation. Posible PNe are marked by: *. 
           The third column lists the spectral-type that we 
           have adopted for each CSPN.} 
\end{table*}


\begin{table*}
\caption[]{Measurements of the strongest stellar emission lines used 
           to classify our sample of [WC] and wels CSPN.}
\label{emiss1}
\centering
\setlength{\tabcolsep}{0.5mm}
\begin{tabular}{lcc|ccc|ccc|c|c}
\hline\hline\noalign{\smallskip}
 name & PN G  & Sp. Type &  \multicolumn{3}{c}{Equivalent Width [\AA]} \vline &  \multicolumn{3}{c}{FWHM [\AA]} \vline & I(4650) & I(4686)   \\
  &   &  & 4650 & 5696 (\ion{C}{iii}) & 5806 (\ion{C}{iv}) & 4650 & 5696 (\ion{C}{iii}) & 5806 (\ion{C}{iv}) & [I(5806) = 100] &  [H$\beta$ = 100] \\
\noalign{\smallskip}\hline\noalign{\smallskip}
 IC 4634  & 000.3+12.2 & wels  & 6 C   & -   & 3    & 22   & -   & 23(d) & 340 & 0.4    \\
 NGC 6790 & 037.8$-$06.3 & wels  & 23 N  & -   & 2    & 18   & -   & 19(d) & $>$1000 &  5.0    \\
 Y-C 2-5  & 240.3+07.0 & wels  & 2 N   & -   & 3    & 19   & -   & 21(d) & 80 &  55.0    \\
 M 4-2    & 248.8$-$08.5 & wels  & 29 N & - & 7:   & 30   & -   & 18:    & 550 &  85.0    \\
 M 3-6    & 253.9+05.7 & wels  & 5 N   & -   & 7    & 25   & -   & 20    & 206 &  3.6    \\
 IC 2501  & 281.0$-$05.6 & wels  & 10 N  & -   & 8    & 22   & -   & 23    & 180 &  -        \\
 IC 2621  & 291.6$-$04.8 & wels  & 93 N  & -   & 5  & 15   & -   & 18    & $>$1000  & 39.0   \\
 He 2-97  & 307.2$-$09.0 & wels  & 15 C  & -   & 8  & 22   & -   & 19    & 290 &  2.0    \\
 NGC 5307 & 312.3+10.5 & wels  & 4 N   & -   & 2  & 16   & -   & 16    & 570 &  43.0    \\
 NGC 5979 & 322.5$-$05.2 & wels  & 17 N & - & 5: & 20   & -   & 18:   & 590 &  107.0    \\
 PC 17    & 343.5$-$07.8 & wels  & 6 N   & -   & +  & 22   & -   & +     & - &  2.0    \\
 IC 4699  & 348.0$-$13.8 & wels  & 8 N  & -  & 2: & 23   & -   & 12:   & 450 &  25.0    \\
 NGC 6337 & 349.3$-$01.1 & wels  & 9: N  & -   & 5  & 50:  & -   & 19    & 210 &  -   \\
\hline
 M 1-53   & 015.4$-$04.5 & wels? & + N   & -   & +    & +    & -   & +     & - &  3.6     \\
 PB 2     & 263.0$-$05.5 & wels? & 11 N  & -   & -    & 16   & -   & -     & - &  12.5    \\
 PB 4     & 275.0$-$04.1 & wels? & 50 N  & -   & -    & 17   & -   & -     & - &  24.4     \\
 IC 2553  & 285.4$-$05.3 & wels? & 64 N & - & ?    & 14   & -   & ?     & - &  34.5     \\
 He 2-128 & 325.8+04.5 & wels? & 7 C   & -   & -  & 21   & -   & -     & - &  0.6    \\
 Cn 1-3   & 345.0$-$04.9 & wels? & 8 C   & -   & -  & 23   & -   & -     & - &  $<$0.1   \\
 IC 4663  & 346.2$-$08.2 & wels? & 48 N  & -   & ?  & 17   & -   & ?     & - &  98.6    \\
 H 1-35   & 355.7$-$03.5 & wels? & 10 C  & -   & ?  & 14   & -   & ?     & - &  $<$0.1   \\
\hline
 H 1-62   & 000.0$-$06.8 & [WC10-11] & 6 N   & 3   & +    & 25   & 10  & +   & $>$1000  & -   \\
 M 1-6    & 211.2$-$03.5 & [WC10-11]? & 4 C   & 4:  & -    & 12   & 29: & -   & -        & -   \\
 M 1-11   & 232.8$-$04.7 & [WC10-11] & 5 N   & 4   & -    & 29   & 10  & -   & -        & -   \\
 M 1-12   & 235.3$-$03.9 & [WC10-11] & 4: C  & 3   & -    & 35:  & 9   & -    & -       & -  \\
 He 2-47  & 285.6$-$02.7 & [WC10-11] & 11 C  & 5 & -  & 28   & 16  & -     & - &  1.1    \\
 He 2-107  & 312.6$-$01.8 & [WC10-11] & 13 N  & 2   &  - & 24   & 13  & -     & - &  5.1     \\
 PHR1416-5809  & 313.9+02.8 & [WC9]     & 160 C & 238 & 82 &   16 &  20 &  25   & $>$160  & - \\
\hline
\end{tabular}
\tablefoot{The emission line at 4650\AA\ is due to C and N. We use the 
same notation as in TA93 to indicate the dominant ion (fourth column).
Also the ``+'' means that the emission is visible, but its measurement is 
uncertain; (d) means that it is possible to distinguish two components, 
ie. \ion{C}{iv} 5801 and 5812\AA.}
\end{table*}


\begin{table*}
\caption{Spectrophotometric standard stars observed
         to flux calibrate each of our spectra. }
\label{std}
\begin{tabular}{lcccc}
\hline\hline\noalign{\smallskip}
 object &    runs   &  STD 1 & STD 2  & STD 3    \\
\noalign{\smallskip}\hline\noalign{\smallskip}
IC 5148      & Nov.  2005 & LTT 377 & LTT 3218 & LTT 9239    \\
He 2-107     & Mar.  2006 & LTT 3864 & CD-32.9927 & -  \\
He 2-47      & Mar.  2006 & LTT 3864 & CD-32.9927 & -  \\
Te 2022      & mar.  2006 & LTT 3864 & - & -  \\
M 1-14       & Mar.  2006 & LTT 3864 & - & -  \\
IC 2621      & Mar.  2006 & LTT 3864 & - & -  \\
M 1-11       & Mar. 2006 & LTT 3864 & - & -  \\
Sp 3         & Aug.  2006 & LTT 7379 & LTT 7987 & LTT 377  \\
He 2-434     & Aug.  2006 & LTT 7379 & LTT 9491 & -  \\
DeHt 1       & Sep.  2006 & LTT 9239 & LTT 1020 & -  \\
A 14         & Sep.  2006 & LTT 9239 & LTT 9491 & -  \\
M 1-53       & Sep.  2006 & LTT 9239 & - & -  \\
SaSt 2-3     & Nov.  2006 & LTT 1020 & - & -  \\
M 4-2        & Nov.  2006 & EG 21 & - & -  \\
PB 4         & Apr. 2007 & HR 3454 & HR 4468 & HR 5501  \\
He 2-105     & Apr.  2007 & HR 3454 & HR 4468 & HR 5501  \\
PC 12        & Apr.  2007 & HR 3454 & HR 4468 & HR 5501  \\
He 2-97      & Apr.  2007 & HR 3454 & HR 4468 & HR 5501  \\
M 1-12       & Apr.  2007 & HR 3454 & HR 4468 & HR 5501  \\
M 1-6        & Apr.  2007 & HR 3454 & HR 4468 & HR 5501  \\
PB 2         & Apr.  2007 & HR 4468 & HR 5501 & -  \\
NGC 6026     & Apr.  2007 & HR 4468 & HR 5501 & -  \\
Wray 17-75   & Apr.  2007 & HR 4468 & HR 5501 & -  \\
He 2-187     & Apr.  2007 & HR 4468 & HR 5501 & -  \\
Y-C 2-5      & Apr.  2007 & HR 4468 & HR 5501 & -  \\
K 2-2        & Apr.  2007 & HR 3454 & HR 5501 & -  \\
PHR1416-5809 & Apr. 2007 & HR 3454 & HR 5501 & -  \\
NGC 6337     & Apr.  2007 & HR 3454 & HR 5501 & -  \\
M 3-6        & Apr.  2007 & HR 3454 & HR 5501 & -  \\
NGC 5307     & Jul. 2007 & HR 4963 & HR 7950 & -  \\
M 1-27       & Jul.  2007 & HR 4963 & HR 7950 & -  \\
NGC 5979     & Jul.  2007 & HR 4963 & HR 7950 & -  \\
H 1-35       & Aug.  2007 & HR 7950 & HR 9087 & HR 718  \\
IC 4634      & Aug.  2007 & HR 7950 & HR 9087 & HR 718  \\
Cn 1-3       & Aug. 2007 & HR 7950 & HR 9087 & HR 718  \\
H 1-62       & Aug.  2007 & HR 7950 & HR 9087 & -  \\
Sa 1-8       & Aug.  2007 & HR 5501 & HR 7596 & HR 8634  \\
H 1-63       & Aug.  2007 & HR 7950 & - & -  \\
Ap 1-12      & Aug.  2007 & HR 7950 & - & -  \\
He 2-128     & May.  2008 & HR5501 & HR 7596 & HR 7950  \\
IC 2501      & May.  2008 & HR5501 & HR 7596 & HR 7950  \\
IC 4699      & May.  2008 & HR5501 & HR7596 & HR 7950  \\
IC 2553      & May. 2008 & HR5501 & HR7596 & HR 7950  \\
PC 17        & Aug.  2008 & HR 7596 & - & -  \\
NGC 6790     & Aug. 2008 & HR 5501 & HR 7596 & -  \\
IC 4663      & Aug. 2008 & HR 5501 & HR 7596 & -  \\
\hline
\end{tabular}
\end{table*}


\newpage

\end{document}